\documentclass[twocolumn, superscriptaddress, prb, showpacs]{revtex4-1}
\newif\ifGALLEYversion\GALLEYversionfalse

\usepackage{graphicx}
\usepackage{amsmath}
\usepackage{multirow}
\usepackage{color}

\newcommand{\mr}[1]{\multirow{2}{*}{#1}}

\begin{document}

\title{Ab initio energetics and kinetics study of H$_2$ and CH$_4$
in the SI Clathrate Hydrate}

\author{Qi Li}
\affiliation{Department of Physics, Wake Forest University,
Winston-Salem, NC 27109, USA}

\author{Brian Kolb}
\affiliation{Department of Physics, Wake Forest University, 
Winston-Salem, NC 27109, USA}

\author{Guillermo Rom\'an-P\'erez}
\affiliation{Departamento de F\'isica de la Materia Condensada,
Universidad Aut\'onoma de Madrid, Cantoblanco, 28049 Madrid, Spain}

\author{Jos\'e M. Soler}
\affiliation{Departamento de F\'isica de la Materia Condensada,
Universidad Aut\'onoma de Madrid, Cantoblanco, 28049 Madrid, Spain}

\author{Felix Yndurain}
\affiliation{Departamento de F\'isica de la Materia Condensada,
Universidad Aut\'onoma de Madrid, Cantoblanco, 28049 Madrid, Spain}

\author{Lingzhu Kong}
\affiliation{Department of Physics and Astronomy, Rutgers University,
Piscataway, New Jersey 08854, USA}

\author{D. C. Langreth}
\affiliation{Department of Physics and Astronomy, Rutgers University,
Piscataway, New Jersey 08854, USA}

\author{T. Thonhauser}
\email[E-mail: ]{thonhauser@wfu.edu}
\affiliation{Department of Physics, Wake Forest University, 
Winston-Salem, NC 27109, USA}

\date{\today}

\begin{abstract}
We present ab initio results at the density functional theory level for
the energetics and kinetics of H$_2$ and CH$_4$ in the SI clathrate
hydrate. Our results complement a recent article by some of the authors
[G. Rom{\'a}n-P{\'e}rez \emph{et al.}, Phys. Rev. Lett. {\bf 105},
145901 (2010)] in that we show additional results of the energy
landscape of H$_2$ and CH$_4$ in the various cages of the host material,
as well as further results for energy barriers for all possible
diffusion paths of H$_2$ and CH$_4$ through the water framework. We also
report structural data of the low-pressure phase SI and the
higher-pressure phases SII and SH.
\end{abstract}

\maketitle


Clathrate hydrates are crystalline, ice-like structures formed out of
water molecules.\cite{Sloan_08} The water framework creates cavities in
which gas molecules---typically O$_2$, H$_2$, CO$_2$, CH$_4$, Ar, Kr,
Xe---can be trapped, which stabilize the framework. The existence of
clathrates was first documented in 1810 by Sir Humphry Davy, and
clathrates became the subject of intensive studies in the 1930s, when
oil companies became aware that clathrates can block
pipelines.\cite{Mao_07} Nowadays, clathrate hydrates are of particular
interest for two reasons: (i) they are formed naturally at the bottom of
the ocean, where they are often filled with CH$_4$.\cite{Buffett_04}
These deposits mean a tremendous stock pile of energy, while---at the
same time---representing a possible global warming catastrophe if
released uncontrolled into the environment through melting; (ii)
clathrate hydrates can be used to store H$_2$ in its cavities and can be
a viable hydrogen-storage material (albeit with moderate
hydrogen-storage density).\cite{Struzhkin_07} For both cases, an
understanding of the interaction between the guest molecule and the host
framework is crucial for their formation and melting processes, which
are still understood poorly.\cite{Gao_05} In this brief report, we
present results that elucidate this crucial
guest-molecule/host-framework interaction and complement a recent paper
by some of the authors.\cite{Soler_10} We show additional results of the
energy landscape of H$_2$ and CH$_4$ in the various cages of the host
material, and we show further results for energy barriers for all
possible diffusion paths of H$_2$ and CH$_4$ through the water
framework. We also report structural data of the phases SI, SII, and SH.

At low pressure, the methane filled clathrate forms the structure SI,
consisting of two types of cages. The smaller cage is built of water
molecules on the vertices of 12 pentagons with a diameter of
approximately 7.86~\AA,\cite{diameter} and we refer to this
as 5$^{12}$ cage, or alternatively as $D$ cage. The larger cage is built
of 12 pentagons and two hexagons with a diameter of approximately
8.62~\AA, and we call it 5$^{12}$6$^2$ or $T$ cage. The
unitcell has cubic symmetry and consists of two $5^{12}$ and six
$5^{12}6^2$  cages, with a total of 46 water molecules. At 250~MPa, the
structure SI transforms into a new cubic phase SII, consisting of
sixteen $5^{12}$ and eight $5^{12}6^4$ cages, containing 136 water
molecules in its unitcell.\cite{Mao_07} When the pressure is increased
to 600~MPa, the structure undergoes another phase transition to the
hexagonal phase SH.\cite{Mao_07} This phase has a smaller unitcell of
three $5^{12}$, two $4^35^66^3$, and one $5^{12}6^8$ cages, with only 34
water molecules. Very nice graphical representations of the different
cages and structures can be found in Refs.~[\onlinecite{Kirchner_04,
Mao_07, Soler_10, Struzhkin_07}]. While other clathrate-hydrate
structures exist, structure SI, SII, and SH  are the most common
ones.\cite{Mao_07}


Guest molecules such as H$_2$ and CH$_4$ in the cavities of the
clathrate hydrates interact with the water framework through van der
Waals forces. But even the water framework itself, i.e.\ the interaction
of water molecules through hydrogen bonds, has a van der Waals
component.\cite{Kolb_11} To capture these effects, we perform here
density functional theory (DFT) calculations utilizing the truly
non-local vdW-DF functional, which includes van der Waals interactions
seamlessly into DFT.\cite{Langreth_05, Thonhauser_07, Langreth_09} We
implemented vdW-DF using  a very efficient FFT
formulation\cite{Soler_09} into the latest release of \textsc{PWscf},
which is a part of the \textsc{Quantum-Espresso} package.\cite{PWscf}
For our calculations we used ultrasoft pseudopotentials with a kinetic
energy cutoff for wave functions and charge densities of 35 Ry and 280
Ry, respectively. A self-consistency convergence criterion of at least
$1\times 10^{-8}$ Ry was used. All structures were fully optimized with
respect to volume and atom positions, and the force convergence
threshold was at least 10$^{-4}$ Ry/a.u.\ for SI and SH. We have also
performed structural calculations on SII, but---due to the large unit
cell with 136 water molecules, i.e.\ 408 atoms---we used a slightly less
tight force convergence criterium of 5$\times$10$^{-4}$ Ry/a.u. For SI
and SH we used a $2\times 2\times 2$ Monkhorst-Pack
k-mesh,\cite{monkhorst76} while for SII we performed $\Gamma$-point
calculations only.


\begin{table*}
\caption{\label{tab:structure} Calculated and experimental lattice
constants $a$ and $c$ for the SI, SII, and SH clathrate hydrates. In
addition, calculated and experimental average nearest-neighbor and
next-nearest-neighbor distances are given, as well as bond angles.
Standard deviations are provided in square brackets. Experimental values
for the lattice constants are taken from Ref.~[\onlinecite{Kirchner_04}]
for methane-filled cages. Experimental values for the averaged
quantities are calculated from the structures given in the supplemental
materials of Ref.~[\onlinecite{Kirchner_04}]. The
experimental distances $d_{\rm O-H}^{\rm nn}$ ($d_{\rm O-H}^{\rm nnn}$)
seem to be underestimated (overestimated), most likely due to the
difficulty of accurately determining H positions in X-ray experiments.
For SI, neutron scattering experiments suggest $d_{\rm O-H}^{\rm nn}=0.97$ \AA\ and
$d_{\rm O-O}^{\rm nn}=2.755$ \AA.\cite{position}
Also note that there is some variation in the experimental results for
the lattice constants in Refs.~[\onlinecite{Sloan_08, Struzhkin_07,
Kirchner_04}].}
\begin{tabular*}{\textwidth}{@{\extracolsep{\fill}}llccccccr@{}}\hline\hline
& & $a$ [\AA] & $c$ [\AA] & $d_{\rm O-H}^{\rm nn}$ [\AA] & $d_{\rm O-H}^{\rm nnn}$ [\AA] &
$d_{\rm O-O}^{\rm nn}$ [\AA] & $\angle_{\rm H-O-H}$ [$^\circ$]& $\angle_{\rm O-O-O}$ [$^\circ$]\\\hline   
\mr{SI}  & calc. & 11.97  & ---    & 0.994 [0.001] & 1.790 [0.014] & 2.781 [0.013] & 107.1$^\circ$ [1.0] & 108.6$^\circ$ [4.0]\\
         & exp.  & 11.88  & ---    & 0.861 [0.031] & 1.911 [0.022] & 2.761 [0.017] & 109.3$^\circ$ [3.0] & 108.7$^\circ$ [3.7]\\
\mr{SII} & calc. & 17.35  & ---    & 0.994 [0.001] & 1.792 [0.016] & 2.784 [0.016] & 107.1$^\circ$ [0.6] & 109.2$^\circ$ [4.3]\\
         & exp.  & 17.19  & ---    & 0.812 [0.016] & 1.959 [0.025] & 2.768 [0.013] & 109.5$^\circ$ [2.0] & 109.3$^\circ$ [4.0]\\
\mr{SH}  & calc. & 12.32  & 10.01  & 0.994 [0.001] & 1.793 [0.015] & 2.782 [0.011] & 107.2$^\circ$ [0.9] & 108.4$^\circ$ [8.5]\\
         & exp.  & 12.33  &  9.92  & 0.781 [0.040] & 1.955 [0.022] & 2.775 [0.005] & 108.9$^\circ$ [5.1] & 108.4$^\circ$ [8.3]\\
\hline\hline
\end{tabular*}
\end{table*}

The empty cages are experimentally not stable, but they have been shown
to be a good starting point for calculations like ours.\cite{Soler_10}
We have calculated the optimized lattice parameters for the SI, SII, and
SH structures and the results are collected in
Table~\ref{tab:structure}. We have also calculated the structures when
filled with methane (one methane molecule per cage) and filled with
hydrogen (up to four H$_2$ per cage), but the lattice parameters expand
less than 0.1\% upon filling, such that we have used the parameters for
the empty cages henceforth. Overall, our optimized lattice constants
agree well with previous calculations\cite{Soler_10} and
experiment.\cite{Kirchner_04} In Table~\ref{tab:structure} we further
analyze the structure of the host materials by calculating the average
nearest-neighbor and next-nearest-neighbor distances and important bond
angles. In general, the calculated average distances and angles vary
only insignificantly amongst SI, SII, and SH, whereas they show a
slightly larger spread for some experimental values.  As a side-note,
for a single water molecule we calculate $d_{\rm O-H}=0.973$~\AA\ and
$\angle_{\rm H-O-H} = 104.9^\circ$, in good agreement with the
experimental numbers of 0.958~\AA\ and 104.5$^\circ$.\cite{water_exp}
Note that vdW-DF is known to give slightly too large binding
distances.\cite{Thonhauser_06, vdW-DF2} Small deviations are visible in
the distances $d_{\rm O-H}^{\rm nn}$ and $d_{\rm O-H}^{\rm nnn}$, which
in sum mostly cancel to give very good agreement with the experimental
O--O distances. Reference~[\onlinecite{Kirchner_04}] also gives the O--O
distances for all structures explicitly as between 2.725~\AA\ and
2.791~\AA, in remarkable agreement with our calculations. Also, our
calculated angles $\angle_{\rm O-O-O}$ agree very well with experiment.
However, the good agreement between oxygen distances and angles---which
describe the structure as a whole---is closely related to the agreement
for the lattice constants.


\begin{figure}[t]
\includegraphics[width=\columnwidth]{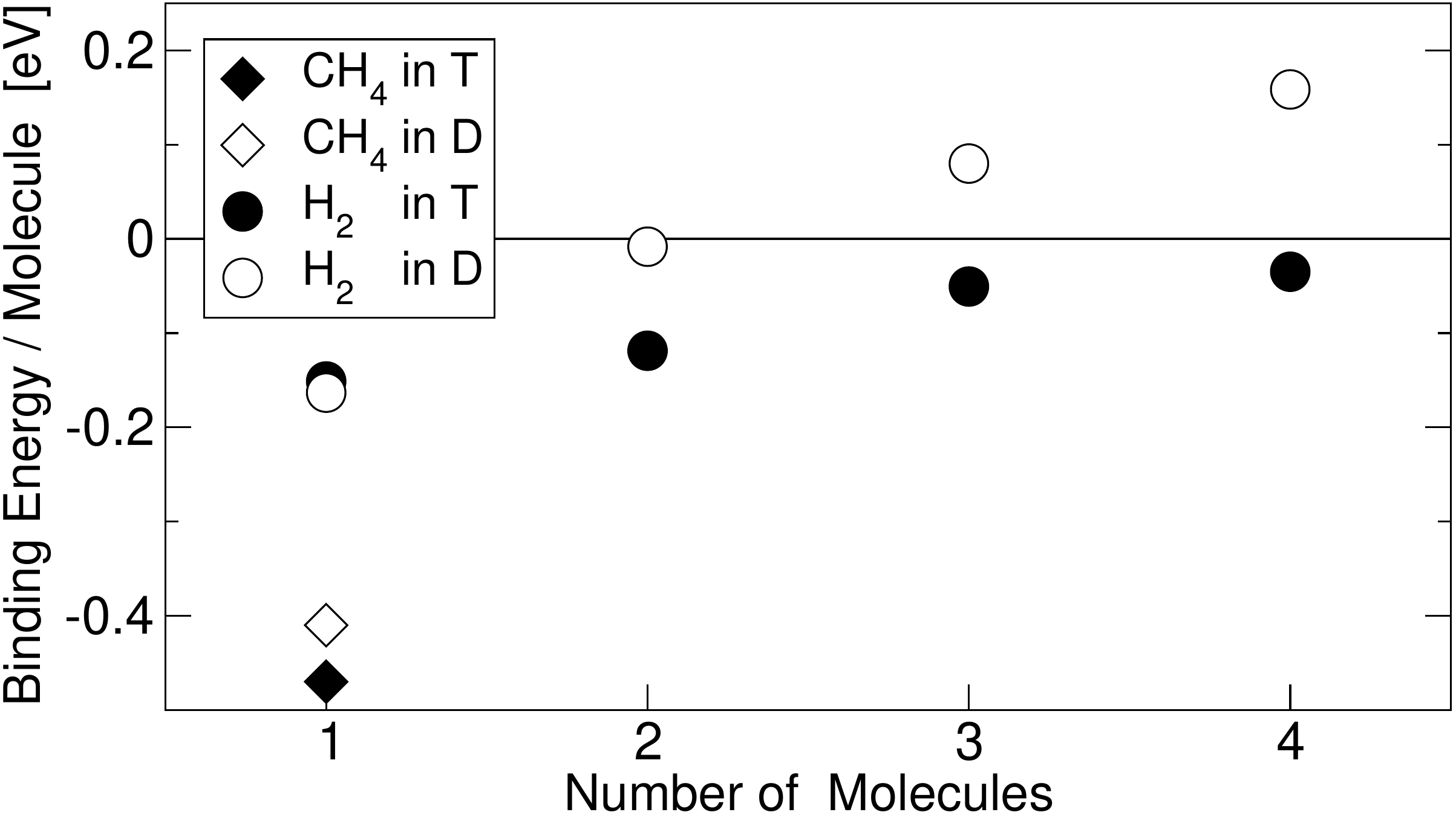}
\caption{\label{fig:E-binding}Binding energies per molecule for different
numbers of CH$_4$ and H$_2$ molecules in the $D$ and $T$ cages of SI.
While the $D$ and $T$ cages can store only one CH$_4$ molecule, the $D$
and $T$ cage can store up to two and four H$_2$ molecules, respectively.}
\end{figure}

We next focus on the binding energies of guest molecules in the SI
structure. In particular, we study the binding energies of CH$_4$ and
H$_2$ in the $D$ and $T$ cages as a function of the number of molecules;
results are depicted in Fig.~\ref{fig:E-binding} for calculations where
molecules are added to only one cage in the unitcell, while all other
cages are kept empty. Here, we define the binding energy as
the energy difference between the ``water-framework + guest-molecules''
system minus the energy of the single constituents. In case of $n$-fold
occupied cages, we subtract $n$ times the energy of the
single molecule. Methane is a large molecule compared to the cage sizes
and it can be seen that in both $D$ and $T$ only one methane molecule
can be stored. Upon adding another methane molecule, the binding energy
increases drastically. The situation is different for the much smaller
H$_2$ molecules. In the smaller $D$ cage we can store up to two H$_2$
molecules, but increasing the number to three or four results in a
positive binding energy, i.e.\ work is required to place more than two
molecules into this cage. Note that the binding energy that we find for
double H$_2$ occupancy is rather small, i.e.\ $-$8 meV/per molecule. It is
thus likely that at non-zero temperatures cages are only singly occupied.
Experimentally, while the majority of recent work seems to favor single
occupancy of the $D$ cages (see e.g.\ Ref.~[\onlinecite{exp1}]), there are
also reports that propose double occupancy
or that find inconclusive evidence.\cite{exp2, exp3, exp4, exp5, exp6}
On the other hand, the larger $T$ cage can
store four H$_2$ molecules. If a fifth molecule is added, it escapes
through one of the hexagonal faces into the neighboring, empty $T$ cage.
Our calculated H$_2$ storage capacity of four molecules in the $T$ cages is in
agreement with experiment.\cite{Struzhkin_07} The binding energy
for one H$_2$ molecule compares well with quantum-chemistry calculations
on isolated cavities, which give --0.123~eV.\cite{exp2} Overall, our
binding energies are slightly smaller than the ones in
Ref.~[\onlinecite{Soler_10}]. Note that quantum motions have been neglected
in our approach, which may play an important role in the binding process
and when determining the cage occupancy. A more precise treatment requires
the computation of the corresponding thermodynamic partition function, as for example
shown in Ref.~[\onlinecite{partition}]. Nevertheless, we consider our
calculations for the binding energy an important first step that already
reveals important information.


\begin{figure}[t]
\includegraphics[height=\columnwidth,angle=-90]{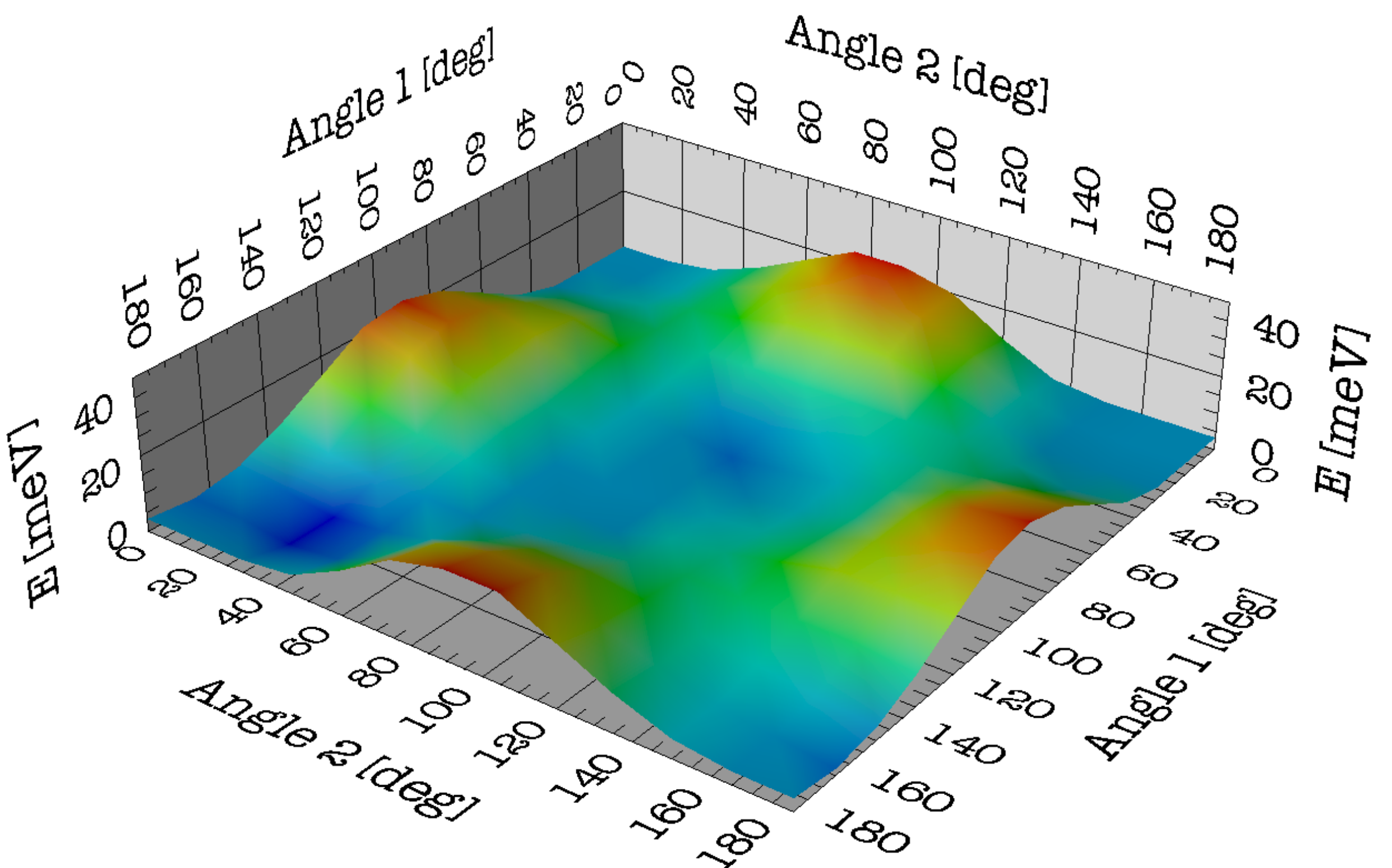}
\caption{\label{fig:D-rotation-energy}Energy landscape in meV for a
rotating methane molecule in a $D$ cage. The $x$ and $y$ axes correspond
to rotations about two mutually perpendicular axis. At the $(0,0)$ point
of the plot, the hydrogen atoms of the methane point exactly toward four
oxygens of the $D$ cage. The difference between the minimum and maximum
energy is 22~meV.}
\end{figure}

It is also interesting to study where and how the H$_2$ and CH$_4$
molecules bind in the cages. If only one molecule is present in the
cages, it binds in the center of the cage. Rotations and small
displacements of H$_2$ in that situation are on an energy scale of
approximately 1~meV and approach the accuracy of our calculations. At
room temperature, such perturbations are thus easily thermally
activated. Since the methane molecule is larger, it cannot move/rotate
as easily. We have studied the rotation of a single methane molecule
centered in the $D$ and $T$ cages as a function of rotation about two
mutually perpendicular axes. The energy landscape for this rotation is
depicted in Fig.~\ref{fig:D-rotation-energy} for the $D$ cage. The $D$
cage with its twelve pentagons and the methane molecule have a related
symmetry, which allows us to choose the $(0,0)$ point of the plot such
that all methane hydrogens point exactly to an oxygen of the host
lattice. At this point, hydrogen bonds are created and the total energy
is the lowest. Upon rotation of the methane, the hydrogen bonds break
and the energy increases. The difference between
the lowest and highest point of this energy landscape is 22~meV,
suggesting thermal activation of rotations at room temperature, and
quantifying an experimental assumption.\cite{Kirchner_04} We have also
studied the rotation of a methane molecule in a $T$ cage and the results
are very similar to the results presented in
Fig.~\ref{fig:D-rotation-energy}, with the difference that the maximal
energy barrier is slightly smaller, i.e.\ 18~meV, not surprising as the
$T$ cage is slightly larger and the methane molecule can rotate more
easily. Calculations for both rotation-energy landscapes
have independently also been performed using
\textsc{Siesta}\cite{Siesta1,Siesta2} and give essentially identical
results.


\begin{figure}[t]
\includegraphics[width=\columnwidth]{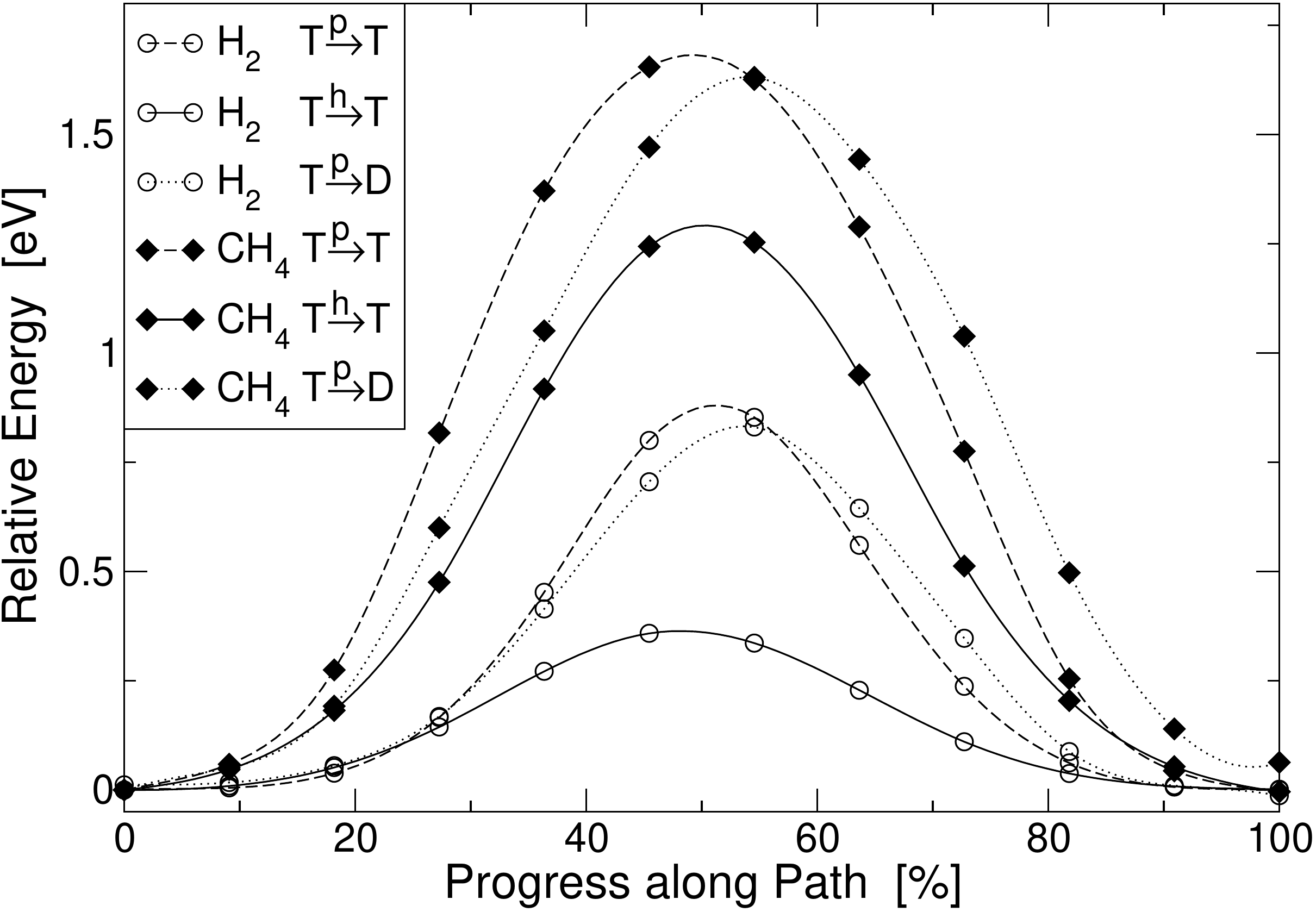}
\caption{\label{fig:NEB}Barriers to diffusion of H$_2$ and CH$_4$
through the water framework along different paths, as a function the
relative progress along the path. The plots are labeled as
$A\stackrel{x}{\to}B$, where $A$ is the type of the starting cage, $B$
is the type of the ending cage, and $x$ refers to either pentagon (p) or
hexagon (h), indicating the opening being used for traversing. The
symbols are the results from 12-image NEB calculations and the lines are
fitted cubic splines, serving as a guide to the eye.}
\end{figure}

Finally, we present results for barriers to diffusion through the water
framework in the SI structure. In the case of H$_2$ and hydrogen storage
this is of much interest, as practical storage solutions require fast
kinetics, i.e.\ low barriers. In the case of CH$_4$, the barriers can
help us understand the natural formation of the filled clathrates. We
have calculated the barriers to diffusion with nudged elastic band (NEB)
calculations, using 12 images along the path from the center of one cage
to the center of the next cage; the results for all possible paths are
plotted in Fig.~\ref{fig:NEB}. Note that the relaxation of the host
lattice is crucial to obtain accurate barrier energies,\cite{Soler_10}
and NEB calculations allow for such relaxations perpendicular to the
path automatically. The plots are labeled as $A\stackrel{x}{\to}B$,
where $A$ is the type of the starting cage, $B$ is the type of the
ending cage, and $x$ refers to either pentagon (p) or hexagon (h),
indicating the opening being used for traversing. Note that for the path
$T\stackrel{p}{\to}D$ there is only one choice of opening, i.e.\ a
pentagon, and the path is not symmetric as the distance from the center
of $T$ to its edge is longer than the corresponding distance in the $D$
cage. Furthermore, this path's end energy is different from its starting
energy, since the guest molecules are binding with different binding
energies in the cages $T$ and $D$, as already evident from
Fig.~\ref{fig:E-binding}. The lowest barriers for H$_2$ and CH$_4$
diffusion agree well with previous calculations.\cite{Soler_10,
Alavi_07} But, our H$_2$ diffusion barrier is an overestimation with
respect to a recent NMR experiment, which gives 0.03~eV and warrants
further investigation.\cite{Okuchi_07} The barriers are in general
smaller for diffusion through hexagons, simply because these openings
are larger.

For hydrogen-storage applications, the low barrier of $\sim$0.3~eV
between $T$ cages (going through a hexagon) is important. Through these
$T$-cage channels, which thread through the material in all three
dimensions, the hydrogen can quickly be absorbed or released. However,
to achieve the material's full storage potential, some hydrogen
molecules will also have to get into the $D$ cages, with a much higher
barrier of $\sim$0.75~eV. The large barrier of $\sim$1.4~eV for methane
diffusion suggests that the methane molecules get trapped while the
clathrate is formed, rather than diffusing into an already existing
empty clathrate.


To conclude, we have performed an ab initio study of structural,
energetic, and kinetic properties of the guest molecules H$_2$ and
CH$_4$ in hydrate clathrates. We have also shown first results for the
difficult-to-model, high-pressure phase SII with a large unit cell,
finding good agreement with experiment.  While we have used vdW-DF for
our study, it is conceivable that its successor, vdW-DF2,\cite{vdW-DF2}
may further improve upon our results. We encourage additional studies of
the hydrate clathrates using vdW-DF2, also including other types of
cages, and more detailed studies of the SII phase, which is one of the
more promising phases amongst the hydrate clathrates for
hydrogen-storage applications. 

We would like to dedicate this report to the memory of
\emph{Prof.\ David Langreth}, who passed away just days before it was
submitted---he is the ``father'' of vdW-DF and his research inspired
many. All calculations were performed on the WFU DEAC cluster.



\begin{thebibliography}{99}
\bibitem{diameter} Here, we define the diameter as twice the average
   distance between oxygen atoms and the cage center.
\end{thebibliography}
\end{document}